# Vanadium dioxide as a natural disordered metamaterial: perfect thermal emission and large broadband negative differential thermal emittance

**Mikhail A. Kats, Romain Blanchard, Shuyan Zhang, Patrice Genevet, Changhyun Ko, Shriram Ramanathan, and Federico Capasso***

School of Engineering and Applied Sciences, Harvard University, Cambridge, Massachusetts 02138, USA

*[capasso@seas.harvard.edu](capasso@seas.harvard.edu)

**We experimentally demonstrate that a thin (~150 nm) film of vanadium dioxide ($VO_2$) deposited on sapphire has an anomalous thermal emittance profile when heated, which arises due to the optical interaction between the film and the substrate when the $VO_2$ is at an intermediate state of its insulator-metal transition (IMT). Within the IMT region, the $VO_2$ film comprises nanoscale islands of metal- and dielectric-phase, and can thus be viewed as a natural, disordered metamaterial. This structure displays "perfect" blackbody-like thermal emissivity over a narrow wavelength range (~40 cm$^{-1}$), surpassing the emissivity of our black soot reference. We observed large broadband negative differential thermal emittance over a >10 °C range: upon heating, the $VO_2$/sapphire structure emitted less thermal radiation and appeared colder on an infrared camera. We anticipate that emissivity engineering with thin film geometries comprising $VO_2$ will find applications in infrared camouflage, thermal regulation, infrared tagging and labeling.**

Thermal radiation is light that is emitted by an object at a temperature above absolute zero. The spectrum and intensity of thermal radiation emitted by an object is a function of its temperature and emissivity, which is in general frequency (*f*) dependent. This is expressed by:

$$I(K,T) = 2hc \frac{K^3}{e^{hcK/k_B T} - 1} \varepsilon(K) \qquad (1)$$

where *I* is the spectral radiance (or the spectral radiant energy density), $K = f/c$ is the wavenumber, *T* is the temperature expressed in Kelvin, *h* is Planck's constant, *c* is the speed of light in vacuum, $k_b$ is the Boltzmann constant, and $\varepsilon(K)$ is the frequency-dependent emissivity [1]. More specifically, the spectral radiance is the radiant power emitted from a unit area of the source per unit solid angle, in the wavenumber interval from *K* to *K* + *dK*, and has units of W cm$^{-1}$. The factor in front of $\varepsilon(K)$ in Eq. 1 is known as Planck's law, and describes blackbody emission. For most objects, $\varepsilon(K)$ is largely independent of temperature (or other external variables such as applied fields).

There is substantial interest in engineering $\varepsilon(K)$ for applications ranging from incandescent light sources [2] to heat management [3] [4] [5] to thermal tagging and imaging [6]. In determining $\varepsilon(K)$ for various materials and structures, frequent use is made of Kirchhoff's law of thermal radiation which states that the emissivity of an object $\varepsilon(K)$ is equal to its frequency-dependent absorptivity $a(K)$ [1].

One approach to engineering $\varepsilon(K)$ has been to select materials with appropriate material dispersion to achieve selective thermal emission [7]. A complementary approach involves surface texturing, either disordered [2] or highly ordered in the case of gratings [8] or photonic crystals [9]. Similarly, photonic cavities can enhance or suppress thermal emission [10] [11]. More recently, optical antennas [12] and metamaterials [13] have also been employed to tailor the directionality and spectrum of thermal emission.

In addition to these static schemes, certain tunable materials can be employed to dynamically manipulate $\varepsilon(K)$. Of particular interest are electrochromic materials such as tungsten oxide ($WO_3$), which undergoes significant change in optical and infrared properties under an applied voltage, and can therefore be used to modulate emissivity of thermal radiators for applications such as temperature control of satellites by radiative cooling [3] [4].

Modulation of the emissivity can also be achieved by using thermochromic materials, whose optical properties are temperature-dependent. Unlike in the case of electrochromic materials, a change in temperature can simultaneously alter the emissivity $\varepsilon(K,T)$ of the object incorporating a thermochromic material, and the blackbody contribution to the spectral radiance $K^3/(e^{hcK/k_BT}-1)$ (see Eqn. 1). A potential benefit of tuning based on thermochromic materials is that it allows for passive "smart" devices that can operate without the need for external power or controls. For example a radiator that has low emissivity at low temperatures and high emissivity at high temperatures can help keep heat in when cold and radiate heat away faster when hot, making it useful for passively maintaining a desired temperature [5].

A commonly studied thermochromic material is vanadium dioxide ($VO_2$), a correlated oxide that experiences a thermally induced insulator-metal transition (IMT) near room temperature ($T_c \sim 67$ °C in bulk crystals), which takes the material from an insulating state to a metallic one. The IMT, which can also be triggered electrically and optically, is the target of research for the realization of a variety of electronic switching devices [14], and finds various uses in optical switching [15]. Present literature on tuning an object's thermal emissivity using $VO_2$ has largely focused on the considerable change of infrared optical properties between the extreme states of the phase transition – fully insulating, and fully metallic [5] [16] [17] [18] [19]. However, very rich physics can be found within the transition region itself, which can be harnessed to obtain additional control over thermal emission properties.

In this letter, we show that a geometry comprising a thin film of $VO_2$ on a sapphire substrate can exhibit "perfect" blackbody-like emissivity (~1) over a narrow range of frequencies when the $VO_2$ is in its transitional state and operates as a natural, tunable metamaterial, i.e. an effective medium with widely-tunable infrared optical properties. As a result of this resonance in emissivity, the sample displays substantial broadband negative differential thermal emittance; i.e. as the sample is heated the thermal emission decreases.

In $VO_2$ thin films, the IMT occurs gradually with increasing temperature: nanoscale inclusions of the metallic phase emerge in the surrounding insulating-phase $VO_2$, which grow and connect in a percolation process, eventually leading to a fully metallic state at the end of the transition [20] [21]. These metallic inclusions are much smaller than the scale of the wavelength at infrared frequencies, and thus $VO_2$ can be viewed as a natural, reconfigurable, disordered metamaterial with variable effective optical properties across the phase transition. In ref. [22], the authors utilized this unique temperature-dependent dispersion of the effective medium to demonstrate that a film of $VO_2$ with thickness much smaller than the wavelength deposited on sapphire can operate as a temperature-tunable absorber; in particular, nearly-perfect absorption was achieved at a particular temperature for a narrow range of infrared wavelengths. The reflectivity of such a device varies dramatically and non-monotonically across the phase transition, with the strong absorption feature appearing during an intermediate state of $VO_2$ as a result of critical coupling to an "ultra-thin-film resonance"; similar resonances have also recently been demonstrated using semiconductor films on metallic substrates in the visible for tailoring reflectivity and absorption [23] [24], and undoped semiconductor films on highly doped semiconductor substrates for tailoring absorption and thermal emission [25]. Since $\varepsilon(K) = a(K)$, such a thin-film $VO_2$ / sapphire structure is expected to have an emissivity ε(K, T) that also depends strongly and non-monotonically on temperature.

Our sample consists of an epitaxial $VO_2$ film of ~150 nm in thickness, grown on a polished single crystal c-plane sapphire by RF-magnetron sputtering with a $V_2O_5$ target (99.9% purity, AJA International Inc.). During film growth, the substrate temperature and RF source gun power were kept constant at 550 °C and 125 W respectively. A mixture of 99.50 sccm Ar and 0.50 sccm $O_2$ was used as the sputtering gas, maintaining the total pressure at 10 mTorr.

We measured the thermal emission from our $VO_2$/sapphire sample by mounting it on a temperature-controlled stage (Bruker A599), changing the temperature from 40 °C to 100 °C and back down (resolution of 0.5 ° in the range of 55 °C – 85 °C and 5 ° outside of that, waiting at least 60 seconds to allow the temperature to settle), and directly sending the emitted light into a nitrogen-purged Fourier transform infrared (FTIR) spectrometer (Bruker Vertex 70) equipped with a room temperature DTGS detector (Fig. 1). As a reference, we replaced the sample with black soot [1], deposited onto a gold-coated silicon wafer using a candle (deposition time > 10 minutes). After deposition, the soot-coated wafer was baked at 200° C for 30 minutes to remove any excess paraffin from the candle. At moderate temperatures, candle-deposited soot is expected to have a wavelength-independent emissivity ε between 0.95 and 0.98 in the infrared [26] [27] [28]; for this work, we assume ε = 0.96. Mid-infrared reflectance measurements taken using a microscope (Bruker Hyperion 2000, NA = 0.4, 15X objective) confirmed that the soot sample is a good blackbody reference (reflectance < 0.01 with some light presumed to be scattered; data not shown).

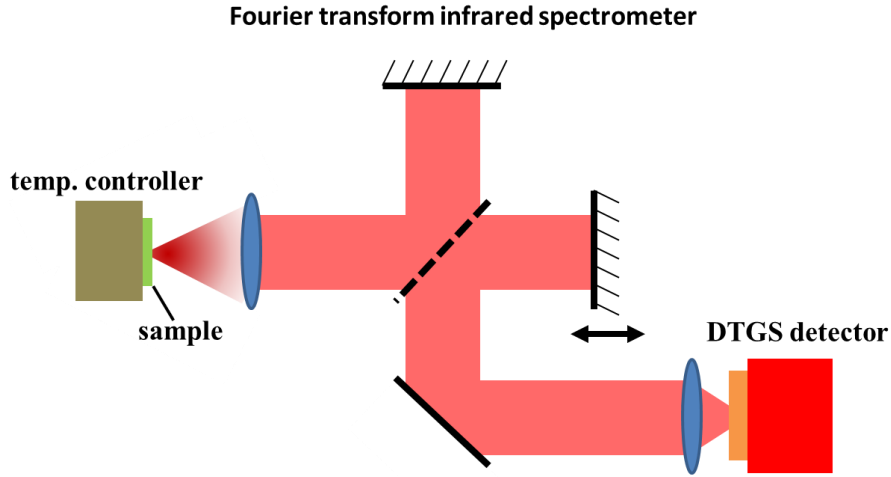

Figure 1. Experimental setup. The $VO_2$/sapphire sample is mounted on a temperature-controlled stage, and the thermal emission is sent into an FTIR spectrometer equipped with a DTGS detector.

To obtain an accurate emission spectrum, we had to correct for the frequency-dependent response of the optics of the FTIR spectrometer and the detector and also account for the thermal emission from sources other than our sample [1] [29]. A particular temperature-dependent measured spectrum $S(K,T)$ can be broken down as

$$S(K,T) = m(K,T)[I(K,T) + b_{instr}(K) - (1 - R(K,T))b_{det}(K)] \qquad (2)$$

where $I(K,T)$ is the actual spectral radiance of the sample, $m(K,T)$ is the instrument response transfer function including the effects of atmospheric absorption along the light path between the sample and the detector, $b_{instr}(K)$ is the thermal contribution of the instrument including emission from the optics and the walls of the FTIR spectrometer (but not from the detector), $b_{det}(K)$ is the thermal emission from the detector, and $R(K,T)$ is the reflectivity of the sample which can be temperature-dependent for thermochromic materials. In our measurement, we assume that $b_{instr}(K) = 0$ because our DTGS detector is at the same temperature as the rest of the instrument, so there is no net flow of thermal radiation between the detector and the optics and walls. There is substantial radiation from the detector itself, so $b_{det}(K)$ cannot be neglected, and a portion of this radiation that enters the interferometer and reaches the sample is reflected back toward the detector; this is accounted for by the $(1 - R(K,T))$ term in Eqn. 2.

Because both our black soot reference and VO$_2$/sapphire sample are opaque within the wavelength range of interest (within the 5-15 μm range, sapphire is opaque as a result of multiple phonon resonances which are present [22] [30]), we can write the emissivity as $\varepsilon(K,T) = a(K,T) = 1 - R(K,T)$, which simplifies Eqn. (2) to

$$S(K,T) = m(K,T)\varepsilon(K,T)[I_{BB}(K,T) - b_{det}(K)] \qquad (3)$$

where $I_{BB}(K,T)$ is the thermal radiation spectrum from a perfect blackbody. Thus, given a reference sample with a known emissivity such as the black soot, one can calculate $m(K,T)$ and $b_{det}(K)$ by measuring the emitted spectrum at two different temperatures $T_1$ and $T_2$, and solving the system of two equations. In our measurement, however, this is unnecessary because $\varepsilon(K,T)$ is factored out in Eqn. (3). Instead we can note that given the measured spectra $S_{sample}(K,T)$ and $S_{ref}(K,T)$ from our sample and reference, respectively, and knowledge of the reference emissivity $\varepsilon_{ref}(K,T)$, we can immediately obtain $\varepsilon_{sample}(K,T)$ by

$$\frac{S_{sample}(K,T)}{S_{ref}(K,T)} = \frac{\varepsilon_{sample}(K,T)}{\varepsilon_{ref}(K,T)}. \qquad (4)$$

Note that it is important to select the appropriate units when representing the Planck distribution function $I_{BB}(K,T)$ as the expressions differ depending on which units are used, e.g. wavenumber or wavelength. Since an FTIR yields spectra with constant resolution in wavenumber [1], we choose to use wavenumber units (which are equivalent to frequency units since $K = f/c$). Furthermore note that the above analysis method is applicable only to samples that are smooth (any roughness must be on a substantially smaller scale than the wavelength of emitted light), which is the case for our VO$_2$/sapphire sample. For rough samples, not all of the light emitted by the detector will be specularly reflected from the sample, and instead some thermal emission from the surrounding area may be scattered into the beam path by the sample; in this case extra care must be taken during data analysis.

We used Eqn. 4 to determine the experimental emissivity of our $VO_2$/sapphire sample, which is plotted for increasing temperatures in Fig. 2(a, b). From the experimental emissivity, we calculated the spectral radiance $I(K,T) = \varepsilon(K,T)I_{BB}(K,T)$ of our black soot reference and the $VO_2$/sapphire sample, shown in Fig. 2(c) for three different temperatures. It can be directly observed that while the thermal emission from the black soot reference is monotonically increasing with increasing temperature, the emission from the $VO_2$/sapphire sample first increases and then decreases. At a particular temperature (~74.5 °C) and wavelength (~864 cm$^{-1}$), the emissivity approaches unity (Fig. 2), indicating that at that wavelength the sample displays nearly "perfect" blackbody-like emission, corresponding to the "perfect absorption" condition demonstrated in ref. [22], which is the result of critical coupling to a strong absorption resonance in the film. The peak in infrared emissivity is relatively broadband (~200 cm$^{-1}$), with the emissivity surpassing that of the black soot reference between 840 cm$^{-1}$ and 885 cm$^{-1}$. The emissivity exhibits hysteresis in temperature (Fig. 1(c)) due to the intrinsic hysteresis in undoped $VO_2$ [14] [22]. Note that the data in Fig. 1(a, b) are shown for increasing temperatures.

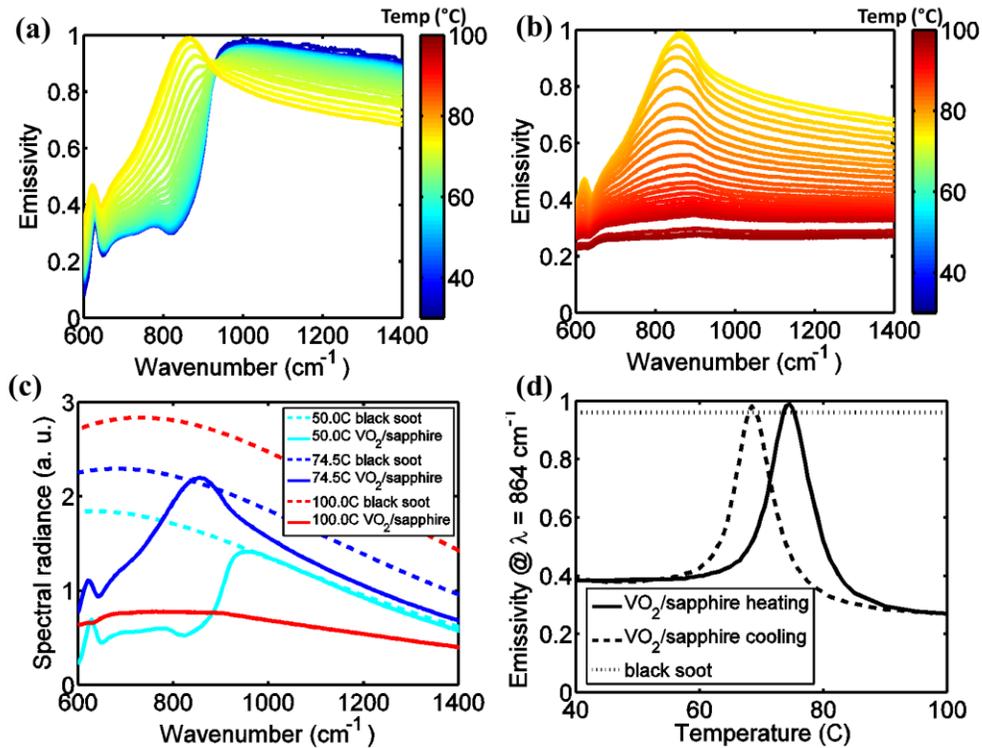

Figure 2. (a, b) Experimentally-determined evolution of the $VO_2$/sapphire emissivity for increasing temperature, separated into ranges of 35 °C – 74.5 °C and 74.5 °C – 100 °C for visual clarity. (c) Thermal emission density (spectral radiance) from black soot (dashed lines) and our $VO_2$/sapphire sample (solid lines) for three different temperatures. The data were taken for increasing temperatures. (d) Thermal emissivity of the $VO_2$/sapphire sample at a wavenumber of 864 cm$^{-1}$ for heating (solid line) and cooling (dashed line), respectively. The dotted line denotes the assumed emissivity of the black soot reference ($\varepsilon = 0.96$).

We integrated $I(K,T)$ of the black soot and $VO_2$/sapphire samples over the 8-14 μm atmospheric transparency window and plotted it as a function of temperature in Fig. 3(a). Plotted this way, it is clear that while heating, the samples displays broadband negative differential thermal emittance over the 73 °C – 85 °C temperature range while heating, and over 68 °C – 80 °C while cooling. The magnitude of the effect is large: over a ~10 °C temperature range the slope is even larger in magnitude than the blackbody slope, indicating that the $VO_2$/sapphire sample has a larger negative differential thermal emittance than the blackbody positive differential thermal emittance over the same temperature range. Imaging these samples with a thermal camera (FLIR Systems Thermovision A40) confirms that due to the negative differential thermal emittance, the sample appears cooler even as it is heating up (Fig. 3(c)). Using the camera, some inhomogeneities in the thermal emittance of the film are visible, which likely result from slight inhomogeneities in the temperature and gradual long-range variations in the film thickness and roughness; these emittance inhomogeneities are amplified around the phase transition temperature.

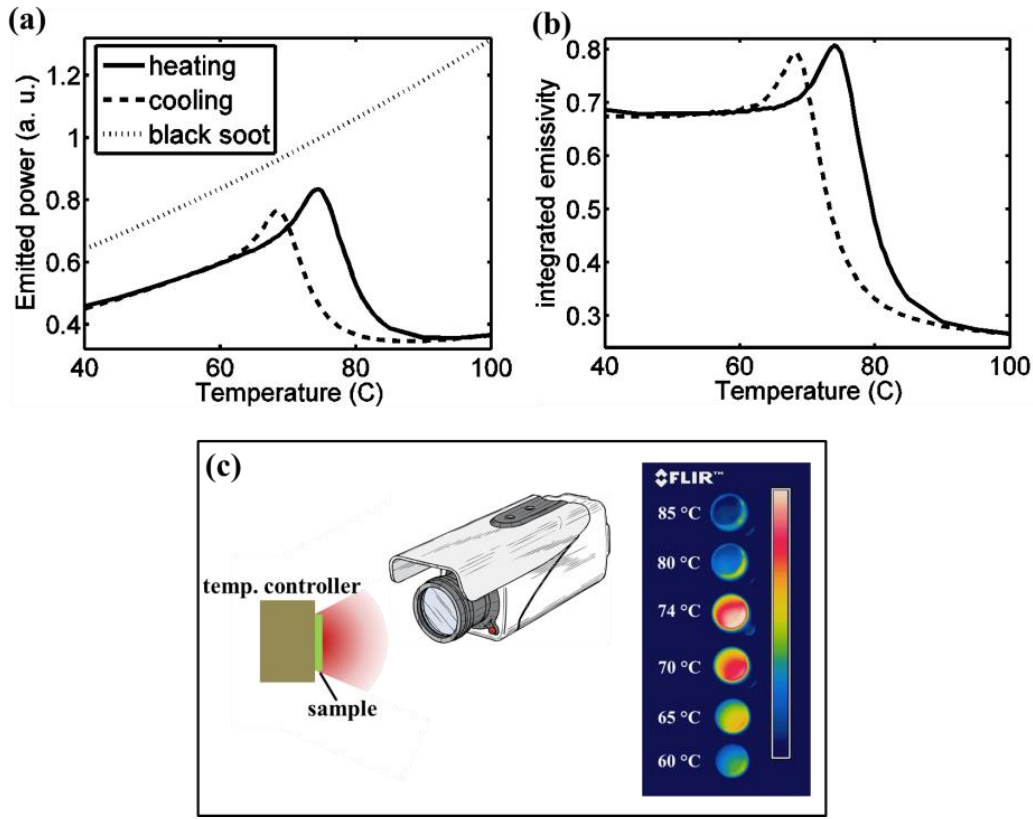

Figure 3. (a) Emitted power of the $VO_2$/sapphire sample integrated over the 8-14 μm atmospheric transmission window for heating (solid line) and cooling (dashed line), compared to the emitted power from black soot. (b) The integrated emissivity of the $VO_2$/sapphire sample over the 8-14 μm wavelength range. (c). Infrared camera images of the sample (diameter = 1 cm) for increasing temperatures.

The $VO_2$/sapphire thin film geometry (and, more generally, any geometries incorporating $VO_2$ with optical resonances in the infrared) is promising for a wide array of applications calling for tunable infrared emissivity, narrow-band "perfect" blackbody-like emission, negative differential thermal emittance, emissivity hysteresis, or some combination thereof. As one example, the emittance profile shown in Fig. 3 can be utilized to make a rewritable infrared "blackboard" by keeping the entire sample at the phase transition temperature, and using a cold or hot probe (such as a laser beam or soldering iron) to "write" messages by locally changing the emissivity. These persistent messages could be viewed with a thermal camera but would otherwise be invisible. A digital version of this device can be used as a rewritable infrared identification tag. As another example, the structure can be used as a type of infrared camouflage: within the ~85 to ~100 ° C temperature region, the total thermally emitted power remains roughly constant, and therefore an infrared camera would not be sensitive to changes in temperature. The width of this flat-emittance region can be extended by decreasing the sharpness of the phase transition, which can be accomplished by introducing defects into the VO$_2$ films [Narayan2006].

Depending on the application, the hysteresis intrinsic to VO$_2$ can either be beneficial (as in memory devices) or detrimental (for devices which require fast on/off switching). Fortunately, a variety of methods to modulate the hysteresis width have been studied, including engineering of the size and shape of grain boundaries [31] and stresses [32], as well as the introduction of various metallic dopants [33] [34]. The aforementioned approaches have also been used to tailor the transition temperature of VO$_2$ within the ~0° C to ~100° C range [31] [32] [33] [14], further expanding the application space.

In conclusion, we have experimentally studied the infrared thermal emittance of a structure comprising a deeply-subwavelength thin film of vanadium dioxide (VO$_2$) on a sapphire substrate. Within the phase transition region of its insulator-metal transition (IMT), the VO$_2$ film comprises nanoscale islands of insulator- and metal-phase VO$_2$ which create a natural, disordered metamaterial with tunable optical dispersion and losses in the infrared, which leads to an absorption resonance within the film that appears and disappears upon temperature tuning. This resonance leads to a large peak in infrared emissivity spanning ~200 cm$^{-1}$, including a 50 cm$^{-1}$ range over which the emissivity of the VO$_2$/sapphire sample is greater than that of black soot, a commonly used blackbody-like emissivity reference. This emissivity peak remains significant even when the emittance spectrum is integrated over the 8-14 μm atmospheric transparency window, and as a result the sample also features a broad-temperature (>10° C) region of which it displays large negative differential thermal emittance such that the sample emits significantly less thermal radiation even as it is heated up. These anomalous emittance properties can find uses in infrared camouflage, thermal regulation, infrared tagging and identification, and other applications.

**Acknowledgements**

We acknowledge helpful discussions with You Zhou, Sergey Shilov, Patrick Rauter, and Yu Yao, and financial support from the AFOSR under grant numbers FA9550-12-1-0289 and FA9550-08-1-0203. M. Kats is supported by the National Science Foundation through a Graduate Research Fellowship. S. Zhang is supported by the Singapore A*STAR National Science Scholarship.

**References**

[1] J. Mink, "Infrared Emission Spectroscopy", Handbook of Vibrational Spectroscopy, DOI: 10.1002/0470027320.s3003 (2006)

[2] A. Y. Vorobyev, V.S. Makin, C. Guo, "Brighter light sources from black metal: significant increase in emission efficiency of incandescent light sources", Physical review letters (2009)

[3] J. S. Hale, M. Devries, B. Dworak, and J. A. Woollam, "Visible and infrared optical constants of electrochromic materials for emissivity modulation applications", Thin Solid Films 313-314, 205-209 (1998)

[4] E. B. Franke, C. L. Trimble, M. Schubert, J. A. Woollam, and J. S. Hale, "All-solid-state electrochromic reflectance device for emittance modulation in the far-infrared spectral region", Applied Physics Letters 77, 930 (2000)

[5] M. Benkahoul, M. Chaker, J. Margot, E. Haddad, R. Kruzelecky, B. Wong, W. Jamroz, and P. Poinas, "Thermochromic VO2 film deposited on Al with tunable thermal emissivity for space applications", Solar Energy Materials and Solar Cells 95, 3504-3508 (2011)

[6] J. K. Coulter, C. F. Kelin, J. C. Jafolla, "Two optical methods for vehicle tagging", SPIE Proceedings 4708 (2002)

[7] D. L. Chubb, A. T. Pal, M. O. Patton, and P. B. Jenkins, "Rare earth doped high temperature ceramic selective emitters", Journal of the European Chemical Society 19, 2551-2562 (1999)

[8] J. J. Greffet, R. Carminati, K. Joulain, J.-P. Mulet, S. Mainguy and Y. Chen, "Coherent emission of light by thermal sources", Nature 416, 61 (2002)

[9] M. U. Pralle, N. Moelders, M. P. McNeal, I. Puscasu, A. C. Greenwald, J. T. Daly, E. A. Johnson, T. George, D. S. Choi, I. El-Kady, and R. Biswas, "Photonic crystal enhanced narrow-band infrared emitters", Applied Physics Letters 81, 4685 (2002)

[10] N. J. Harrick and A. F. Turnet, "A thin film optical cavity to induce absorption or thermal emission", Applied Optics 9, 2111 (1970)

[11] I. Celanovic, D. Perreault, and J. Kassakian, "Resonant-cavity enhanced thermal emission", Physical Review B 72, 075127 (2005)

[12] J. A. Schuller, T. Taubner, and M. L. Brongersma, "Optical antenna thermal emitters", Nature Photonics 3, 658 (2009)

[13] X. Liu, T. Tyler, T. Starr, A. F. Starr, N. M. Jokerst, and W. J. Padilla, "Taming the blackbody with infrared metamaterials as selective thermal emitters", Physical Review Letters 107, 045901 (2011)

[14] Z. Yang, C. Ko, and S. Ramanathan, Annual Review of Materials Research 41, 337 (2011)

[15] W. R. Roach and I. Balberg, Solid State Communication 9, 551 (1971)

[16] F. Guinneton, L. Sauques, J. C. Valmalette, F. Cros, J. R. Gavarri, "Comparative study between nanocrystalline powder and thin film of vanadium dioxide VO2: electrical and infrared properties", Journal of Physics and Chemistry of Solids 62, 1229-1238 (2001)

[17] F. Guinneton, L. Sauques, J.-C. Valmalette, F. Cros, J.-R. Gavarri, "Optimized infrared switching properties in thermochromic vanadium dioxide thin films: role of deposition process and microstructure", Thin Solid Films 446, 287-295 (2004)

[18] R. L. Voti, M. C. Larciprete, G. Leahu, C. Sibilia, and M. Bertolotti, "Optimization of thermochromic VO2 based structures with tunable thermal emissivity", Journal of Applied Physics 112, 034305 (2012)

[19] A. J. Topping, "Structure with variable emittance", US Patent 0155154 (2004)

[20] H. S. Choi, J. S. Ahn, J. H. Jung, T. W. Noh, and D. H. Kim, Physical Review B 54, 7 (1996)

[21] M. M. Qazilbash, M. Brehm, B.-G. Chae, P.-C. Ho, G. O. Andreev, B.-J. Kim, S. J. Yun, A. V. Balatsky, M. B. Maple, F. Keilmann, H.-T. Kim, and D. N. Basov, Science 318, 1750 (2007)

[22] M. A. Kats, D. Sharma, J. Lin, P. Genevet, R. Blanchard, Z. Yang, M. M. Qazilbash, D. N. Basov, S. Ramanathan, and F. Capasso, "Ultra-thin perfect absorber employing a tunable phase change material", Applied Physics Letters 101, 221101 (2012)

[23] M. A. Kats, R. Blanchard, P. Genevet, and F. Capasso, "Nanometer optical coatings based on strong interference effects in highly absorbing media", Nature Materials 12, 20 (2013)

[24] H. Dotan, O. Kfir, E. Sharlin, O. Blanck, M. Gross, I. Dumchin, G. Ankonina, and A. Rothschild, "Resonant light trapping in ultrathin films for water splitting", Natuer Materials 12, 158-164 (2013)

[25] W. Streyer, S. Law, G. Rooney, T. Jacobs, and D. Wasserman, "Strong absorption and selective emission from engineered metals with dielectric coatings", Optics Express 21, 9113 (2013)

[26] R. J. Lauf, C. Hamby, M. A. Akerman, A. W. Trivelpiece, "Blackbody material", US Patent 5313325 (1994))

[27] N. Gao, H. Sun, D. Ewing, "Heat transfer to impinging round jets with triangular tabs", International Journal of Heat and Mass Transfer 46, 2557–2569 (2003)

[28] M. H. Friedman, "Passive control of emissivity, color and camouflage", US Patent 5734495 (1998)

[29] "Emission Measurements" document from Bruker provided by S. Shilov

[30] F. Gervais and B. Piriou, "Anharmonicity in several-polar-mode crystals: adjusting phonon self-energy of LO and TO modes in Al2O3 and TiO2 to fit infrared reflectivity", J. Phys. C: Solid State Phys 7, 2374 (1974)

[31] J. Narayan and V. M. Bhosle, "Phase transition and critical issues in structure-property correlations of vanadium oxide", Journal of Applied Physics 100, 103524 (2006)

[32] Y. Muraoka, Y. Ueda, Z. Hiroi, Large modification of the metal-insulator transition temperature in strained VO2 films grown on TiO2 substrates, "Journal of Physics and Chemistry of Solids 63, 965-967 (2002)

[33] H. Futaki and M. Aoki, "Effects of various doping elements on the transition temperature of vanadium oxide semiconductors", Japanese Journal of Applied Physics

[34] M. Nishikawa, T. Nakajima, T. Kumagai, T. Okutani, and T. Ssuchiya, "Adjustment of thermal hysteresis in epitaxial VO2 films by doping metal ions", Journal of the Ceramic Society of Japan 119, 577-580 (2011)